\documentstyle[epsfig]{aipproc}

\begin{document}
\title{Cosmological Reionization\footnote{Invited Paper,
Plenary Session, to appear in 
{\it Proceedings of the 20$^{\rm th}$ Texas
Symposium on Relativistic Astrophysics and Cosmology}, eds.
H. Martel and J. C. Wheeler (AIP Conference Series), in press (2001).}}

\author{Paul R. Shapiro}
\address{Dept. of Astronomy, The University of Texas, Austin, TX 78712 USA}

\maketitle

\begin{abstract}
The universe was reionized by redshift $z\approx6$ by a small fraction
of the baryons in the universe, which released energy following their
condensation out of a cold, dark, and neutral IGM
into the earliest galaxies. The theory of this 
reionization is a critical missing link in the theory of galaxy formation. 
Its numerous observable consequences include effects on the spectrum, 
anisotropy and polarization of the 
cosmic microwave background and signatures of high-redshift star and quasar 
formation.
This energy release also created feedback on galaxy formation
which left its imprint on the mass spectrum and internal 
characteristics of galaxies and on the gas between galaxies long
after reionization was complete. Recent work suggests that the 
photoevaporation of dwarf galaxy minihalos may have consumed most of the 
photons required to reionize the currently-favored $\Lambda$CDM universe.
We will review recent developments in our understanding of
this process.
\end{abstract}

\section{Introduction}

Observations of quasar absorption spectra indicate that the universe was reionized prior to redshift $z=5$ (e.g. \cite{shcm99}). 
CMB anisotropy data on the first acoustic peak set limits
on the electron scattering optical depth of
a reionized IGM which imply $z_{\rm rei}\lesssim40$
(model-dependent) \cite{gbl99}. 
Together with the
lack of a hydrogen Ly$\alpha$ resonance scattering (``Gunn-Peterson'')
trough in the spectrum of an SDSS quasar discovered
at $z=5.8$ \cite{fanetal00},
these limits currently suggest that $6\lesssim z_{\rm rei}\lesssim 40$.
The origin and consequences of this
reionization are among the major unsolved problems of cosmology.
(For prior reviews and further references, see, e.g. 
\cite{bl01,hk99,shapiro95}.) 
Photons emitted by
hitherto undetected massive stars or miniquasars formed within
early galaxies are generally thought to be the reionization source.

This is consistent with the theoretical expectation in cosmological models
like the Cold Dark Matter (CDM) model, in which the first objects to
condense out of the background and possibly begin star formation
were small, of subgalactic mass (i.e. $M\lesssim10^6M_\odot$) and began
to form as early as $z\gtrsim30$
(e.g. \cite{go97,hl97,sgb94,tegmarketal97}). 
The initial collapse of such objects
only led to star formation if radiative cooling 
was possible after collapse, usually involving $\rm H_2$ molecules,
and it is likely 
that the first objects to form stars released radiation (and possibly
SN explosion energy as well) which exerted a strong feedback on the
subsequent formation history of other objects. 
The details of this feedback and even the overall {\it sign} (i.e. 
negative or positive) are poorly understood
(e.g. 
\cite{bvm92,cfa98,cr86,ferrara98,har99,hrl96,hrl97,ki00,shapiro95,sgb94,sm95}).
It appears that the first stars would have photodissociated $\rm H_2$ long
before enough UV was emitted to bring about reionization, so it is
currently thought that if reionization was accomplished by stars,
they formed in objects with virial temperature
$T_{\rm vir}>10^4\rm K$, of mass
$\gtrsim10^8M_\odot$, which were able to cool by atomic radiative cooling,
even without $\rm H_2$. This conclusion changes, 
however, if the first sources were
miniquasars whose nonthermal spectra had a significant X-ray flux, since
this would have created a {\it positive} feedback on the
$\rm H_2$.

If we adopt an optimum efficiency for massive star formation and
radiation release by the collapsed baryon fraction in a standard CDM model
which is flat, matter-dominated, and COBE-normalized
(an unrealistic
model which has too high an amplitude to satisfy X-ray cluster
abundance constraints at $z=0$, but conservatively overestimates
$z_{\rm rei}$), $z_{\rm rei}\approx50$ is possible
\cite{shapiro95}. But more suitable CDM
models all tend to yield $z_{\rm rei}\lesssim20$ 
(e.g. \cite{bnsl01,co00,cfgj00,go97,hl97,sgb94,vs99}).

\section{Cosmological Ionization Fronts}

{\bf Ionization Fronts in a Clumpy Universe.}
The neutral, opaque IGM
out of which the first bound objects condensed was dramatically reheated
and reionized at some time between a redshift $z\approx50$ and $z\approx6$
by the radiation released by some of these objects.
When the first sources turned on, they
ionized their surroundings by propagating weak, R-type
ionization fronts which moved outward supersonically with respect to both
the neutral gas ahead of and the ionized gas behind the front, racing ahead
of the hydrodynamical response of the IGM \cite{shapiro86a,sg87}. 
The
problem of the time-varying radius of a spherical I-front which surrounds 
isolated sources in a cosmologically-expanding IGM 
was solved analytically by \cite{shapiro86a,sg87}, taking
proper account of the I-front jump condition generalized to cosmological
conditions. They applied these solutions to determine when 
and how fast these I-front-bounded spheres
would grow to overlap and, thereby,
complete the reionization of the universe.
The effect of density inhomogeneity on the rate of I-front propagation
was described by a mean ``clumping factor'' $c_l>1$, which
slowed the I-fronts by increasing the average recombination rate per H atom
inside clumps. This suffices to describe the average
rate of I-front propagation as long as the
clumps are either not self-shielding or, if so, only
absorb a fraction of the ionizing photons emitted by the central source. 
In two recent calculations, this analytical prescription was adapted
to N-body simulations of structure formation in the CDM model, to
calculate the evolving size of the spherical H~II 
regions with which to surround
putative sources of ionizing radiation, to model the growth of the ionized 
volume filling factor leading to reionization \cite{bnsl01,cfgj00}.

Numerical radiative transfer methods are currently under
development to solve this problem in 3D for the inhomogeneous density 
distribution which arises as cosmic structure forms, but so far 
are limited to an imposed 
density field without gas dynamics (e.g. \cite{anm99,cfmr00,nus01,rs99}). 
A different approach, which replaces radiative transfer with a ``local optical
depth approximation,''
intended to mimic the average rate at which I-fronts expanded and 
overlapped during reionization within the context of cosmological gas 
dynamics simulation, has also been developed \cite{gnedin00a}.
These recent attempts to model inhomogeneous reionization
numerically are handicapped by their limited spatial resolution
($\gtrsim1\,\rm kpc$), which prevents them from resolving the
most important density inhomogeneities.
The dynamical response of density inhomogeneities to the I-fronts which
encountered them and the effect which
these inhomogeneities had, in turn, on the
progress of universal reionization, therefore, require further analysis. 
Toward this end, we have developed a radiation-hydrodynamics code
which incorporates radiative transfer and have focused our attention
on properly resolving this small-scale structure.
In what follows, we summarize
the results of new radiation-hydrodynamical simulations of 
what happens when a
cosmological I-front overtakes a gravitationally-bound density inhomogeneity
-- a dwarf galaxy minihalo -- during reionization.  
According to \cite{ham00}, the photoevaporation of these
sub-kpc-sized objects is likely to be
the dominant process by which ionizing photons were
absorbed during reionization, so this problem is of critical importance
in determining how reionization proceeded. 

{\bf Dwarf Galaxy Minihalos at High Redshift.} The effect which small-scale
clumpiness had on reionization depended upon the sizes, densities, and spatial
distribution of the clumps overtaken by the I-fronts during reionization.
For the currently-favored $\Lambda$CDM model $(\Omega_0=1-\lambda_0=0.3$, 
$h=0.7$, $\Omega_bh^2=0.02$,
primordial power spectrum index $n_p=1$; COBE-normalized),
the universe  at $z>6$
was already filled with dwarf galaxies capable of 
trapping a piece of the global, intergalactic I-fronts which reionized the
universe and photoevaporating their gaseous baryons back into the IGM
(see Figure~1). Prior to their encounter with these I-fronts,``minihalos''
with $T_{\rm vir}<10^4\rm K$ were neutral and 
optically thick to hydrogen
ionizing radiation, as long as their total mass exceeded
the Jeans mass $M_J$ in the unperturbed background IGM prior to reionization
[i.e. $M_J=5.7\times10^3(\Omega_0h^2/0.15)^{-1/2}
(\Omega_bh^2/0.02)^{-3/5}((1+z)/10)^{3/2}M_\odot$], as was required to
enable baryons to collapse into the halo along with dark matter. Their
``Str\"omgren numbers'' $L_S\equiv2R_{\rm halo}/\ell_S$, the ratio of a halo's 
diameter to its Str\"omgren length $\ell_S$ inside the halo (the length
of a column of gas within which the unshielded arrival rate of ionizing
photons just balances the total recombination rate), were large.
For a uniform gas of H density $n_{H,c}$, located a distance $r_{\rm Mpc}$
(in Mpc) from a UV source emmitting $N_{\rm ph,56}$ ionizing photons (in units
of $10^{56}\rm s^{-1}$), the Str\"omgren length is only 
$\ell_S\approx(100\,{\rm pc})(N_{\rm ph,56}/r_{\rm Mpc}^2)
(n_{H,c}/0.1\,{\rm cm}^{-3})^{-2}$, so $L_S\gg1$ for a wide range of
halo masses and sources of interest.
In that case, the intergalactic,
weak, R-type I-front which entered each minihalo during reionization would
have decelerated to about twice the sound speed of the ionized gas before
it could exit the other side, thereby transforming itself into a D-type
front, preceded by a shock. Typically, the side facing the source would
then have expelled a supersonic wind backwards toward the source, which
shocked the surrounding IGM as the minihalo photoevaporated.

The importance of this photoevaporation process has long been
recognized in the study of interstellar clouds exposed to ionizing
starlight (e.g. \cite{bertoldi89,bm90,ksw83,ll94,lcgh96,mellemaetal97,os55,swk82,spitzer78}).
In the cosmological context, however,
its importance has only recently been fully appreciated. 

\begin{figure}[b!] 
\centerline{\epsfig{file=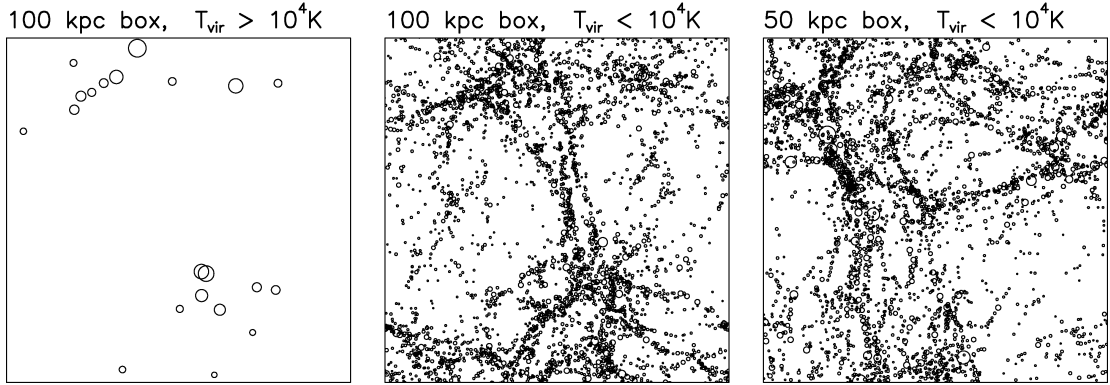,height=2.1in}}
\vspace{10pt}
\caption{{\bf\char'003 CDM Halos at High Redshift:}
DM Halos at $z=9$ are shown with sizes and locations
determined by FOF algorithm applied 
to $\rm P^3M$ simulations ($128^3$ particles, $256^3$ cells),
projected onto face of
simulation cube of proper size $L_{\rm box}$, as labelled. (a) 
(left) 
$M_{\rm halo}>10^{7.6}M_\odot$ only (i.e. $T_{\rm vir}>10^4K$). (b)
(middle) $10^{5.6}M_\odot<M_{\rm halo}<10^{7.6}M_\odot$ only
(i.e. $T_{\rm vir}<10^4K$) -- THE MINIHALOS;
(c) (right) like (b), but higher resolution simulation
with same number of particles and cells in 1/8 volume (i.e.
$M_{\rm halo,min}=10^{4.7}M_\odot$).
}
\label{fig1}
\end{figure}

In proposing the expanding minihalo model to explain Ly$\alpha$
forest (``LF'') quasar absorption lines, \cite{bss88}
discussed how gas originally confined by the gravity of dark-matter
minihalos in the CDM model would have been expelled by pressure forces
if photoionization by ionizing background radiation suddenly
heated all the gas to an isothermal condition at $T\approx10^4\rm K$,
a correct description only in the optically thin limit.
The first discussion of the photoevaporation of a primordial density
inhomogeneity overtaken by a cosmological I-front, including
radiation-hydrodynamical simulations, however,
was in \cite{srm97,srm98}.
\cite{bl99} subsequently estimated the relative
importance of this process for dwarf galaxy minihalos of different masses
at different epochs
in the CDM model, concluding that 50\%--90\% of 
the gas which had already collapsed into 
gravitationally bound objects when reionization
occurred should have been photoevaporated.

\begin{figure}[b!] 
\centerline{\epsfig{file=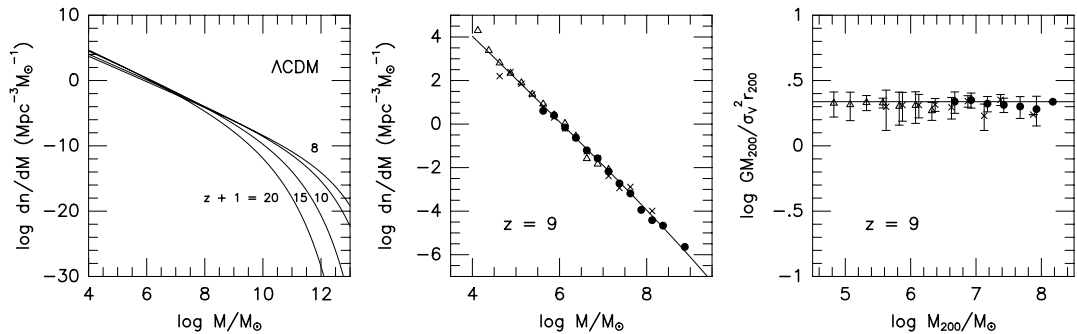,height=2in}}
\vspace{0pt}
\caption{{\bf CDM Halos: N-body Results Vs. PS+TIS Approximations:}
(a) (left) Proper Number Density
of Halos per Unit Halo Mass, $dn_{\rm halo}/dM$,
in $\Lambda$CDM  
at different redshifts, as labelled, according
to PS approximation; (b) (middle) Same as (a), with PS mass function at $z=9$
(solid curve) compared with N-body results
for halos in Fig.~1 for three different resolutions:
$L_{\rm box}=100\,\rm kpc$ (circles), $50\,\rm kpc$ (crosses), and
$25\,\rm kpc$ (triangles); (c) Virial ratio, $GM_{200}/(\sigma_V^2r_{200})$,
versus halo mass $M_{200}$, where $\sigma_V^2$ is the halo DM 
velocity dispersion and $M_{200}$ and $r_{200}$ are the mass and radius
of the sphere whose average density is 200 times the cosmic mean value,
for
all halos in Fig.~1 with at least
200 particles per halo [symbols same as in (b), with
$1\sigma$ error bars]. Horizontal line is analytical prediction of the TIS 
model.}
\label{fig2}
\end{figure}

Not only did this photoevaporation during reionization affect most
of the collapsed baryons, but so common were minihalos with
$T_{\rm vir}<10^4K$ during reionization that the typical reionizing photon
is likely to have encountered one of them, according to \cite{ham00}. 
A catalogue of all halos
in a cubic volume 100 kpc (50 kpc) on a side at $z=9$ in the currently-favored
$\Lambda$CDM model based on N-body simulations reveals
that minihalos with $T_{\rm vir}<10^4K$ (i.e. $M<10^{7.6}M_\odot$)
were separated on average by only $d\approx7\,\rm kpc$ (3.5 kpc) for
$M\geq10^{5.6}M_\odot$ ($M\geq10^{4.7}M_\odot$), respectively,
while their geometric cross sections together covered $f\sim16\%$ (30\%) of
the area along every 100 kpc of an average line of sight \cite{smi01}
[see Figs.~1(b), (c)]. If the
sources of reionization, on the other hand, were larger-mass halos with 
$T_{\rm vir}>10^4K$ [like those in Fig.~1(a)], then
these were well-enough separated that typical reionization photons were 
likely to have been absorbed by
intervening photoevaporating minihalos.
To demonstrate this in a statistically meaningful way with more
dynamic range than the N-body results can yet provide, an analytical 
approximation to the detailed numerical results is required. We shall
combine the well-known Press-Schechter (PS) prescription for deriving the
average number density of halos of different mass at each epoch with the
nonsingular truncated isothermal sphere (TIS) model of
\cite{is01,sir99} for the size, density profile, and temperature of each halo
as unique functions of the halo mass and collapse redshift for a given 
background universe. (The TIS model and further tests and applications
of it are briefly mentioned elsewhere in this volume \cite{is01b}).
We illustrate the validity of these aproximations in Figure~2 by
comparing them
with the N-body simulation results depicted in Figure~1. The PS halo mass
functions plotted in Figure~2(a) are shown in Figure~2(b) to reproduce
the N-body results over a range of $10^5$ in halo mass, for which
$dn_{\rm halo}/dM\approx10^{12}(M/M_\odot)^{-2}{\rm Mpc}^{-3}M_\odot$
(proper units) at $z=9$. Likewise, Figure~2(c) shows that the TIS
model correctly predicts the average virial ratio, $[GM_{200}/(\sigma_V^2r_{200})]_{\rm TIS}=2.176$, for halos of different
mass according to the N-body results, over this same mass range.

\begin{figure}[b!] 
\vspace{-10pt}
\centerline{\epsfig{file=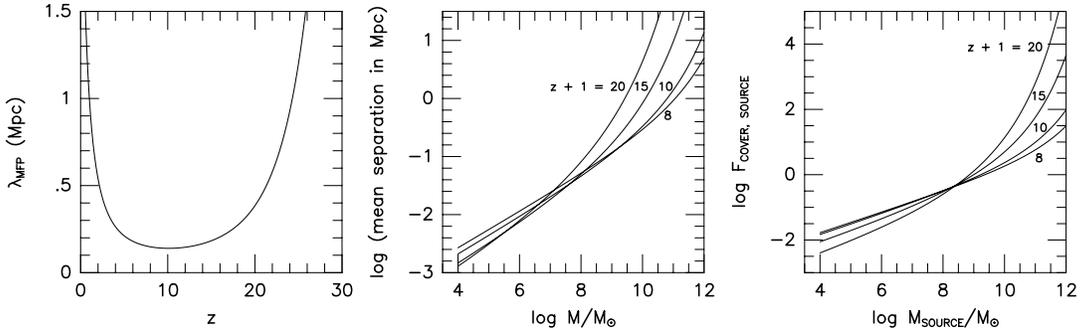,height=2in}}
\vspace{-5pt}
\caption{{\bf Minihalo Sinks and Sources:}
(a) (left) Proper mean free path $<\!n_{\rm halo}\sigma_{\rm halo}\!>^{-1}$
for absorption of photons by minihalos in $\Lambda$CDM at different 
$z$ (i.e. if they photoevaporate at $z_{\rm ev}<z$); (b) (middle)
Proper mean 
separation of halos $<\!d_{\rm sep}\!>\equiv(M\,dn_{\rm halo}/dM)^{-1/3}$
versus halo mass at different redshifts in $\Lambda$CDM, as labelled;
(c) (right) Fraction of sky covered by minihalos located within the
mean volume per source halo, 
$F_{\rm cover,source}=<\!d_{\rm sep}\!>/\lambda_{\rm mfp}$, versus source halo 
mass, at different redshifts.}
\label{fig3}
\end{figure}

This TIS+PS approximation allows us to determine
which halo masses are subject to photoevaporation and how common they are,
as functions of redshift. Each minihalo with $T_{\rm vir}<10^4{\rm K}$
is opaque to H-ionizing photons with a geometric cross section 
$\sigma_{\rm halo}=\pi r_t^2$, where 
$r_t=\hbox{TIS radius}\approx(0.75\,{\rm kpc})(M/10^7M_\odot)^{1/3}
[(1+z_{\rm coll})/10]^{-1}(\Omega_0h^2/0.15)^{-1/3}$. Ionizing photons
travelling through this universe will suffer absorption by
these minihalos [i.e. those
with mass $M_J\leq M\leq M(10^4K)$ at each $z$]
with a mean free path $\lambda_{\rm mfp}$, as shown in Figure~3(a).
For comparison,
the mean separation $<\!d_{\rm sep}\!>$ of halos of each mass is plotted
in Figure~3(b). At $z=9$, for example, $\lambda_{\rm mfp}=160\,{\rm kpc}$,
while halos of mass $M\gtrsim10^8M_\odot$ are separated on average
by $<\!d_{\rm sep}\!>\approx50(M/10^8M_\odot)^{1/3}\rm kpc$.
As shown in Figure~3(c), their ratio,
$<\!d_{\rm sep}\!>/\lambda_{\rm mfp}$, gives the fraction of the sky, as 
seen by a source halo of a given mass, which is covered by
opaque minihalos located within the mean volume per source halo.
If halos with $M\gtrsim10^8M_\odot$ are the reionization sources, 
their minihalo covering fraction is close to unity and increases with
increasing source mass. This estimate will 
increase by a factor of a few if we take account of the statistical
bias by which minihalos tend to cluster around the source halos \cite{ham00}.

An argument like this led
\cite{ham00} to argue that our photoevaporating 
minihalos are the chief consumers of the ionizing photons responsible for
reionization. As a result, they suggest, the photoevaporation of these
minihalos drives the number of ionizing photons per baryon required to
reionize the universe up by an order of
magnitude compared to previous estimates!
A recent semi-analytical study of inhomogeneous reionization by
\cite{mhr00}, which neglected this
effect, concluded that only one ionizing photon per hydrogen
atom would have been sufficient to reionize most of the volume of the IGM
by $z\approx5$, an order of magnitude too low when compared to the new
photoevaporation-dominated estimates. The gas clumping model they adopted
apparently missed the smallest scales because it was adjusted
to match numerical simulation results which could not resolve these
scales. Simulations by \cite{gnedin00a},
which agreed with \cite{mhr00}, have the same problem
since their
resolution limit exceeded $\sim1\,\rm kpc$, too large to resolve
the minihalos which photoevaporate.

\begin{figure}[b!] 
\vspace{-10pt}
\centerline{\epsfig{file=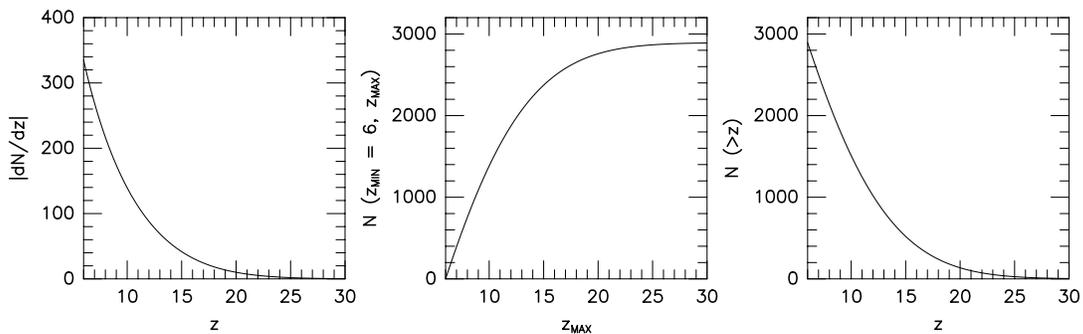,height=2in}}
\vspace{-5pt}
\caption{{\bf Minihalos Encountered By a Photon Emitted At High Redshift:}
(a) (left) Number of minihalos per unit redshift interval in $\Lambda$CDM
encountered by a photon at redshift $z$ which travels along the average LOS;
(b) (middle) Total number
of minihalos in $\Lambda$CDM along photon path which probes redshift interval 
$z_{\min}\leq z\leq z_{\max}$ for $z_{\min}=6$ is plotted
versus $z_{\max}$; (c) (right) Asymptotic number of
halos along photon path which probes redshift interval $(z,\infty)$.}
\label{fig4}
\end{figure}

{\it We are led to conclude that further study of the photoevaporation of
cosmological minihalos during reionization is essential if we
are to advance the theory of reionization.} The new 
``photoevaporation-dominated'' reionization scenario suggests that 
significantly more
than one photon per baryon may have been required, and this is
difficult to understand on the basis of simple extrapolation 
(e.g. \cite{gs96,mhr99}) of
either quasar or stellar photon production rates observed at
$z<5$ to $z>5$. Perhaps this problem will be alleviated by appeal to the 
recently revised stellar output for zero metallicity stars \cite{bkl00,ts00},
or to a
higher escape fraction of ionizing photons from their source
than the nominal $f_{\rm esc}\approx0.1$
typically assumed (e.g. \cite{dsf00,leithereretal95,rs00,spa01,wl00}).
Alternatively, if minihalo sources, alone, reionized the IGM (e.g. with
miniquasars instead of starlight), then their ionizing photons
could have done so without encountering any other minihalos, since
$F_{\rm cover,source}<1$ for halos with $T_{\rm vir}<10^4\rm K$.
Photons released, thereafter, would have encountered fewer opaque minihalos,
as well,
since baryons would not then
have condensed out of a reheated IGM to form {\it new}
minihalos, according to \cite{sgb94}.
 
\begin{figure}[b!] 
\vspace{-40pt}
\centerline{\epsfig{file=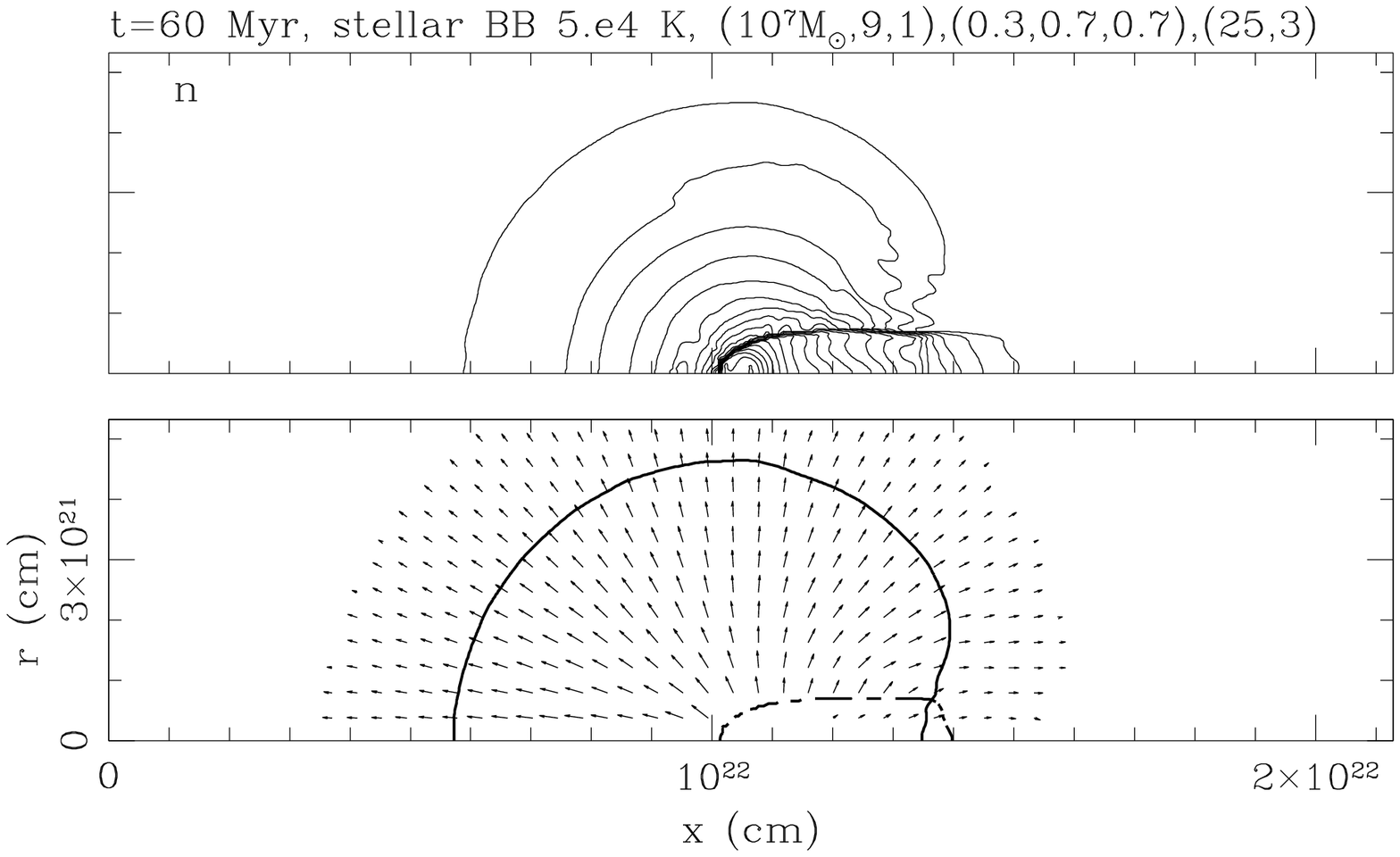,width=3.1in}\hskip-0.2in
            \epsfig{file=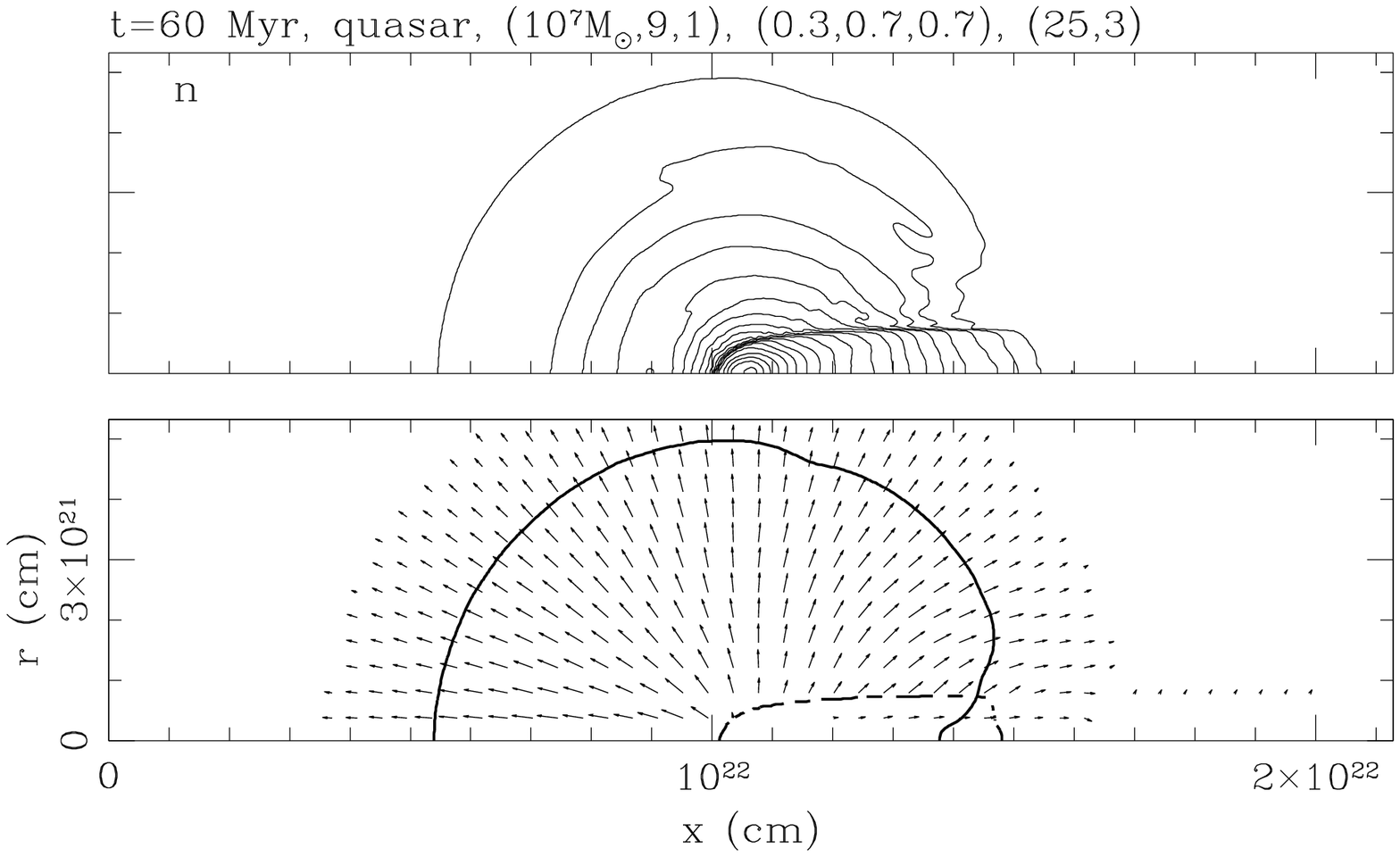,width=3.1in}}
\vspace{-50pt}
\caption{{\bf Photoevaporating Minihalo I.}
One time-slice, 60 Myr after I-front caused by 
ionizing source (located far to the left of
computational box along the $x$-axis) overtakes a $10^7M_\odot$
minihalo [centered at $(r,x)=(0,1.06\times10^{22}{\rm cm})$]
at $z=9$ in the $\Lambda$CDM universe, for two types
of source spectra:
(a) (left) STELLAR CASE and (b) (right) QUASAR CASE.
(upper panels) isocontours of atomic density, 
logarithmically spaced, 
in $(r,x)-$plane of cylindrical coordinates; (lower panels)
velocity arrows are plotted with length proportional to gas velocity.
An arrow of length equal to the spacing between arrows has velocity
$25\, {\rm km\, s^{-1}}$; minimum velocities plotted are
$\rm 3\,km\,s^{-1}$. Solid line shows current extent of gas
initially inside minihalo at $z=9$.
Dashed line is I-front (50\% H-ionization contour).}
\label{fig5}
\end{figure}

The good news is that observations should be able to distinguish these possibilities. In order for photoevaporating 
minihalos to have consumed most
of the ionizing photons produced before reionization is complete, the
covering fraction of the ionization sources by these
photoevaporative flows must be of order unity. Hence,
if we can observe the sources of 
universal reionization directly, we will generally be able to observe
foreground photoevaporating minihalos in absorption toward these
sources. Such observations will constrain and help diagnose
the reionization process.
As shown in Figure~4, photons emitted by {\it any} high $z$ source before or
during reionization will typically encounter large numbers of
photoevaporating minihalos at $z>6$. The number of minihalos probed per
unit redshift interval, in fact, is just $(1+z)^{-1}$ times the ratio of
the horizon size $c/H(t)$ at $z$ to $\lambda_{\rm mfp}$, a large number. It
is important, therefore, for us to predict the effect which
reionization will have on the gas in these minihalos.

\begin{figure}[b!] 
\centerline{\epsfig{file=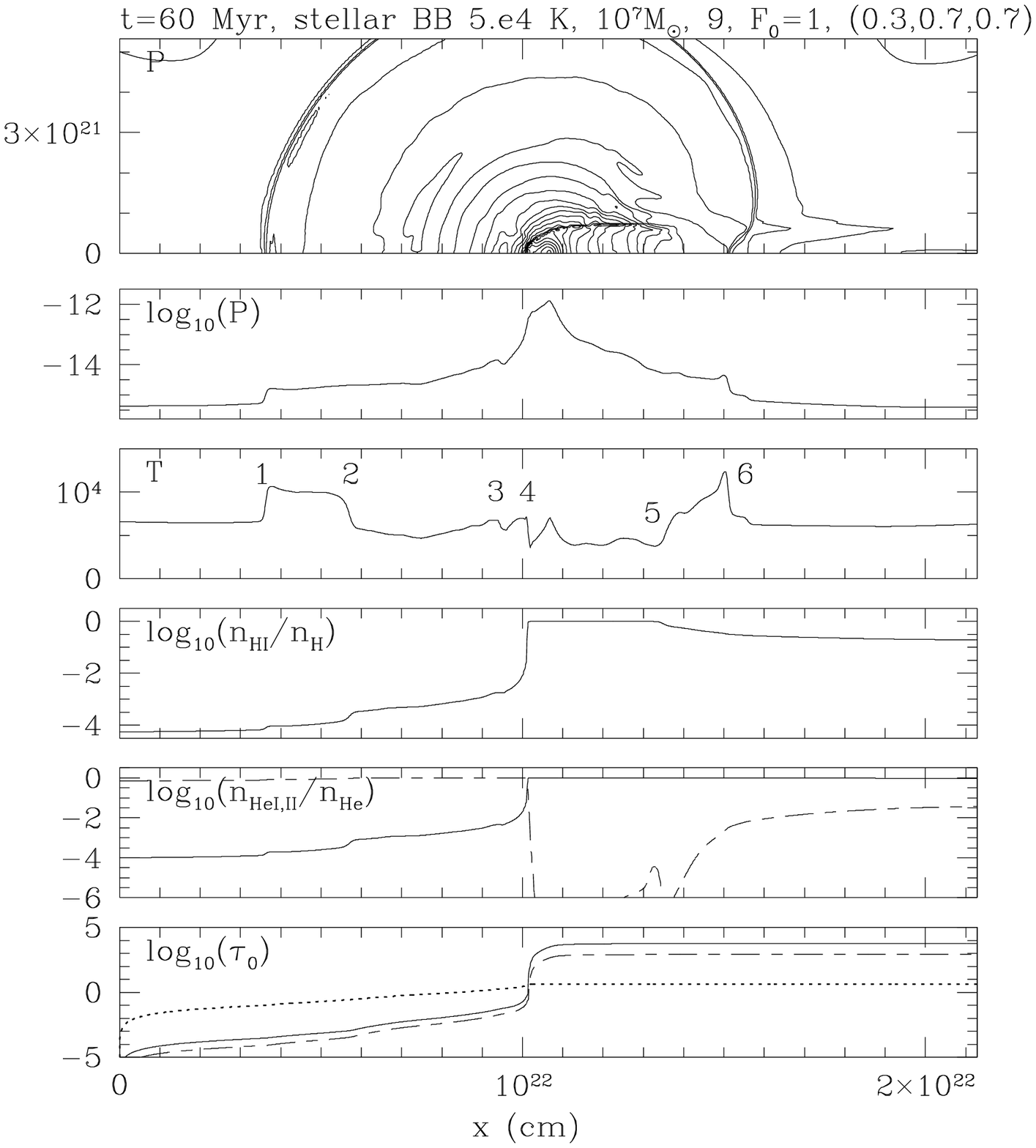,width=3.4in}\hskip-0.5in
            \epsfig{file=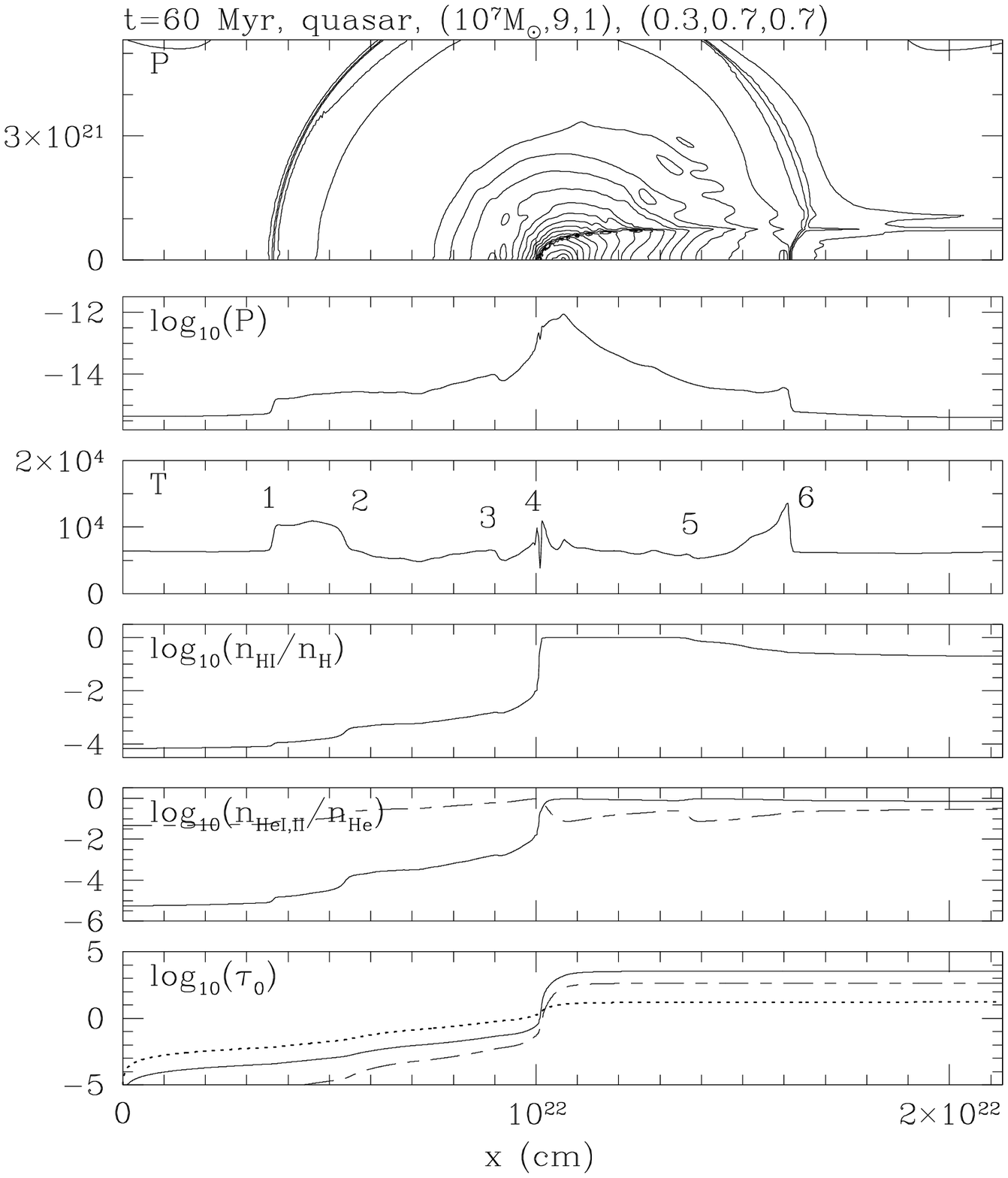,width=3.4in}}
\vspace{10pt}
\caption{{\bf Photoevaporating Minihalo II.}
Same time-slice as Fig. 5.
(a) (left) STELLAR CASE and (b) (right) QUASAR
CASE. From top to bottom: (i)
isocontours of pressure, logarithmically spaced, in
$(r,x)-$plane of cylindrical coordinates; (ii) pressure along the
$r=0$ symmetry axis; (iii) temperature; (iv) H~I fraction; (v) He~I
(solid) and He~II (dashed) fractions; (vi) bound-free optical
depth measured from $x=0$ along the $x$-axis,
at the threshold ionization energies for
H~I (solid), He~I (dashed), He~II (dotted). 
Key features of the flow are indicated by the numbers which label them on 
the temperature plots: \hbox{1 = IGM} shock; \hbox{2 = contact}
discontinuity which separates
shocked halo wind (between 2 and 3) from swept-up IGM (between 1 and 2); 
\hbox{3 = wind}
shock; between 3 and \hbox{4 = supersonic} wind; \hbox{4 = I-front}; 
\hbox{5 = boundary} of gas initially inside minihalo at $z=9$; 
6 = shock in shadow region caused by compression of shadow gas 
by shock-heated gas outside shadow.}
\label{fig6}
\end{figure}

{\bf The Photoevaporation of Dwarf Galaxy Minihalos Overtaken
by Cosmological Ionization Fronts.}
We have performed radiation-hydrodynamical simulations
of the photoevaporation of a cosmological minihalo overrun by a weak,
R-type I-front in the surrounding IGM, created by an external source 
of ionizing radiation \cite{sr00a,sr00b,sr00c,sri01}. The gas contained
H, He, and a possible heavy 
element abundance of $10^{-3}$ times solar.
Our simulations in 2D, axisymmetry used an
Eulerian hydro code with Adaptive Mesh Refinement and
the Van~Leer flux-splitting algorithm, which
solved nonequilibrium ionization rate equations (for H, He, C, N, O, Ne,
and S) and included an explicit treatment of radiative transfer
by taking into account the bound-free opacity of H and He
\cite{mellemaetal97,rml97,rtcb95}.
The reader is referred to \cite{srm97,srm98,srm00}
for earlier results which considered uniform clouds and
demonstrated the importance of a proper treatment of
optical depth.

\begin{figure}[b!] 
\vspace{-10pt}
\centerline{\epsfig{file=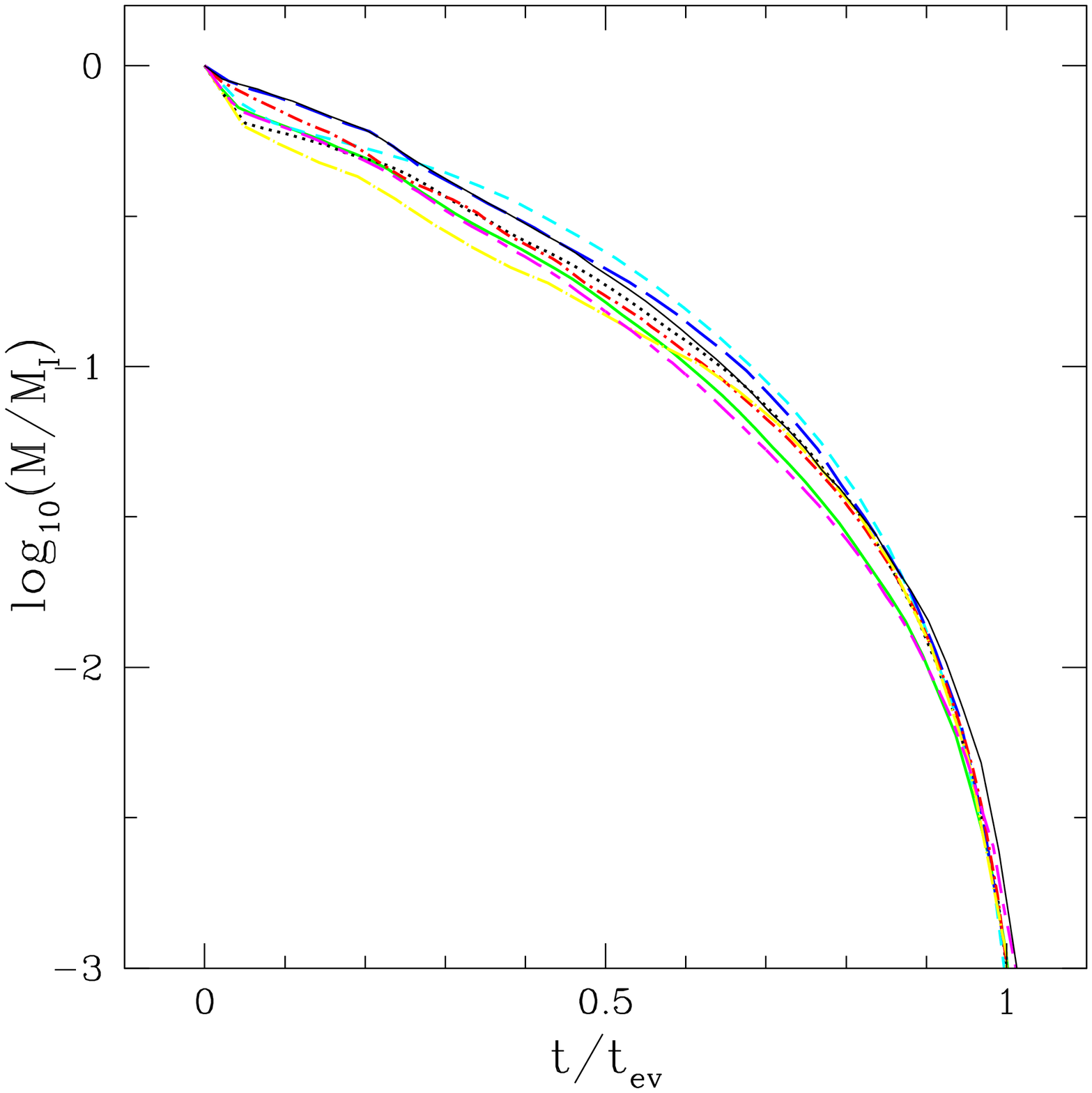,width=2.9in}
            \epsfig{file=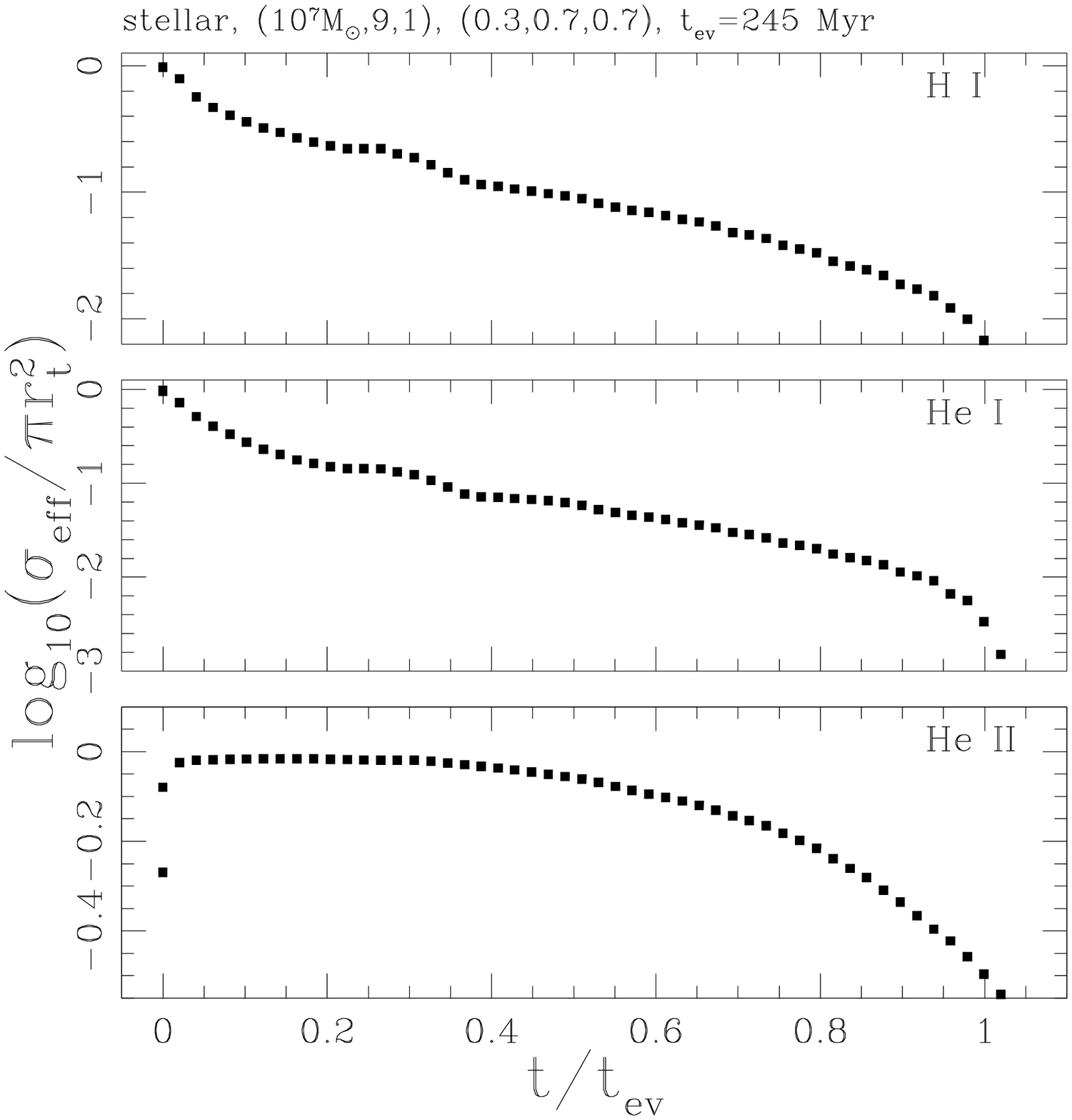,width=2.9in}}
\vspace{10pt}
\caption{{\bf Evolution of Neutral Gas Content of Photoevaporating
Minihalo.} (a)~(left)
Fraction of mass $M_I$, the mass which is initially inside
the minihalo when the intergalactic I-front overtakes it, which remains
neutral versus time $t$ (in units of the evaporation time
$t_{\rm ev}$ at which this $M/M_I=10^{-3}$), for a large range of cases
with different assumed source spectra and fluxes and minihalo masses;
(b) (right) fraction of minihalo initial geometric cross section
$\pi r_t^2$ which is opaque to source photons that can ionize
H~I (top panel), He~I (middle panel), or He~II (bottom panel) versus
time (in units of $t_{\rm ev}$), for the case with stellar source shown
in Figs.~5, 6 ($t_{\rm ev}=245\,{\rm Myr}$).
}
\label{fig7}
\end{figure}

Here we compare some of our results of minihalo photoevaporation
in a $\Lambda$CDM universe
\cite{sri01} for two types of source
spectra: a quasar-like source with emission spectrum 
$F_\nu\propto\nu^{-1.8}$ ($\nu>\nu_H$) and a stellar 
source with a 50,000 K blackbody spectrum, with luminosity 
and distance adjusted to keep the ionizing photon fluxes the 
same in the two cases. In particular, if $r_{\rm Mpc}$ is the proper
distance (in Mpc) between source and minihalo and $N_{{\rm ph},56}$ is the 
H-ionizing photon luminosity (in units of $10^{56}\,s^{-1}$), then the
flux at the location of the minihalo would, if unattenuated,
correspond initially to $N_{{\rm ph},56}/r^2_{\rm Mpc}=1$;
thereafter, $r_{\rm Mpc}\propto a(t)$, the cosmic scale factor.

Our initial condition before ionization is
that of a $10^7M_\odot$ minihalo in the $\Lambda$CDM
universe
which collapsed out and virialized at $z_{\rm coll}=9$,
yielding a truncated, nonsingular isothermal sphere  (``TIS'') of radius 
$r_t=0.75\,\rm kpc$ in hydrostatic equilibrium
with $T_{\rm vir}=4000\,\rm K$
and dark-matter velocity dispersion $\sigma_V=5.2\,\rm km\,s^{-1}$.
The TIS profile
has a central density and an average density which are
18,000 and 130 times the mean background density, respectively, with
core radius $r_0\equiv r_{\rm King}/3\sim r_t/30$.
This hydrostatic sphere is embedded in a self-similar,
spherical, cosmological infall according to \cite{bertschinger85}.

\begin{figure}[b!] 
\vspace{-10pt}
\centerline{\epsfig{file=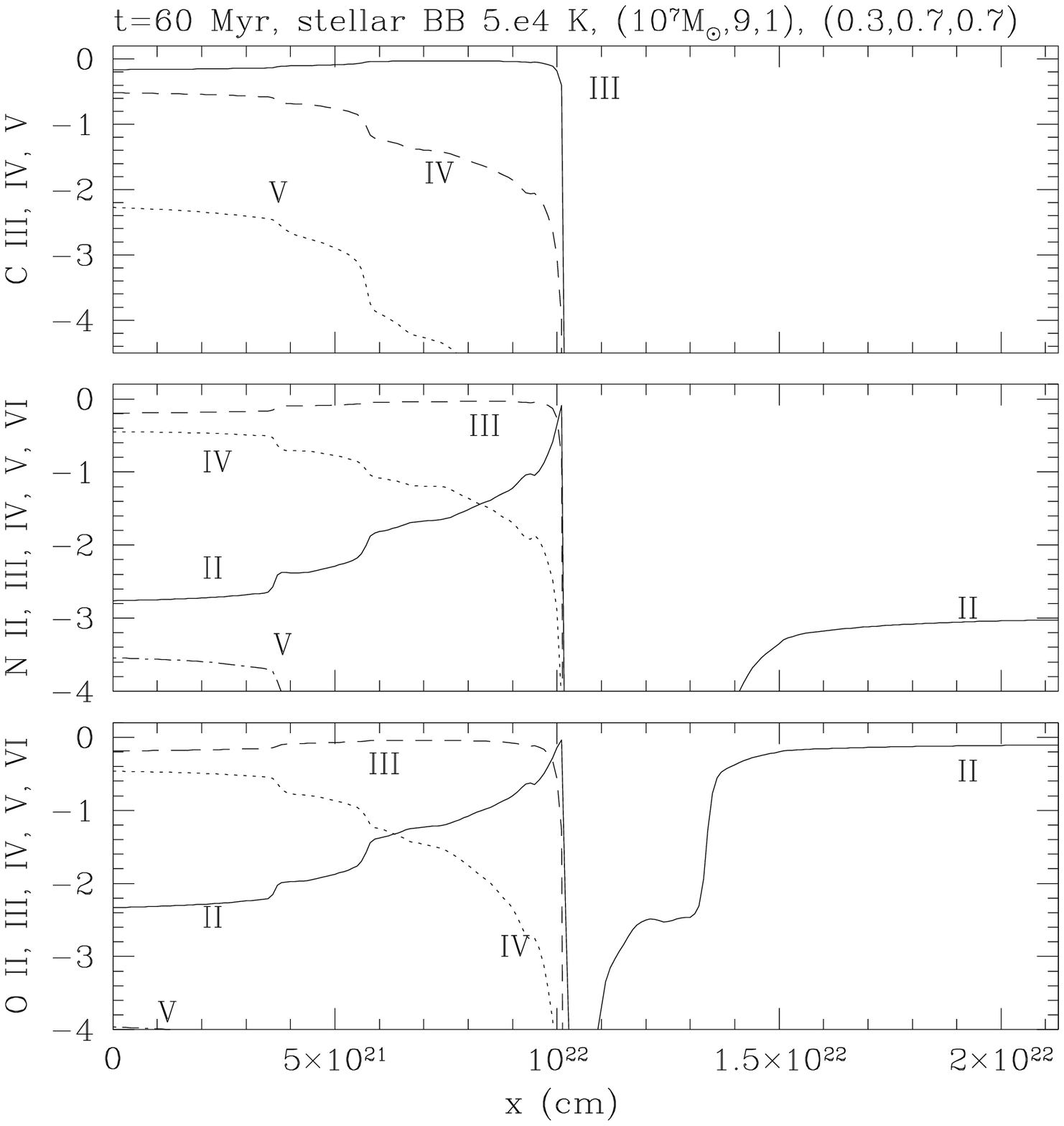,width=3.2in}\hskip-0.4in
            \epsfig{file=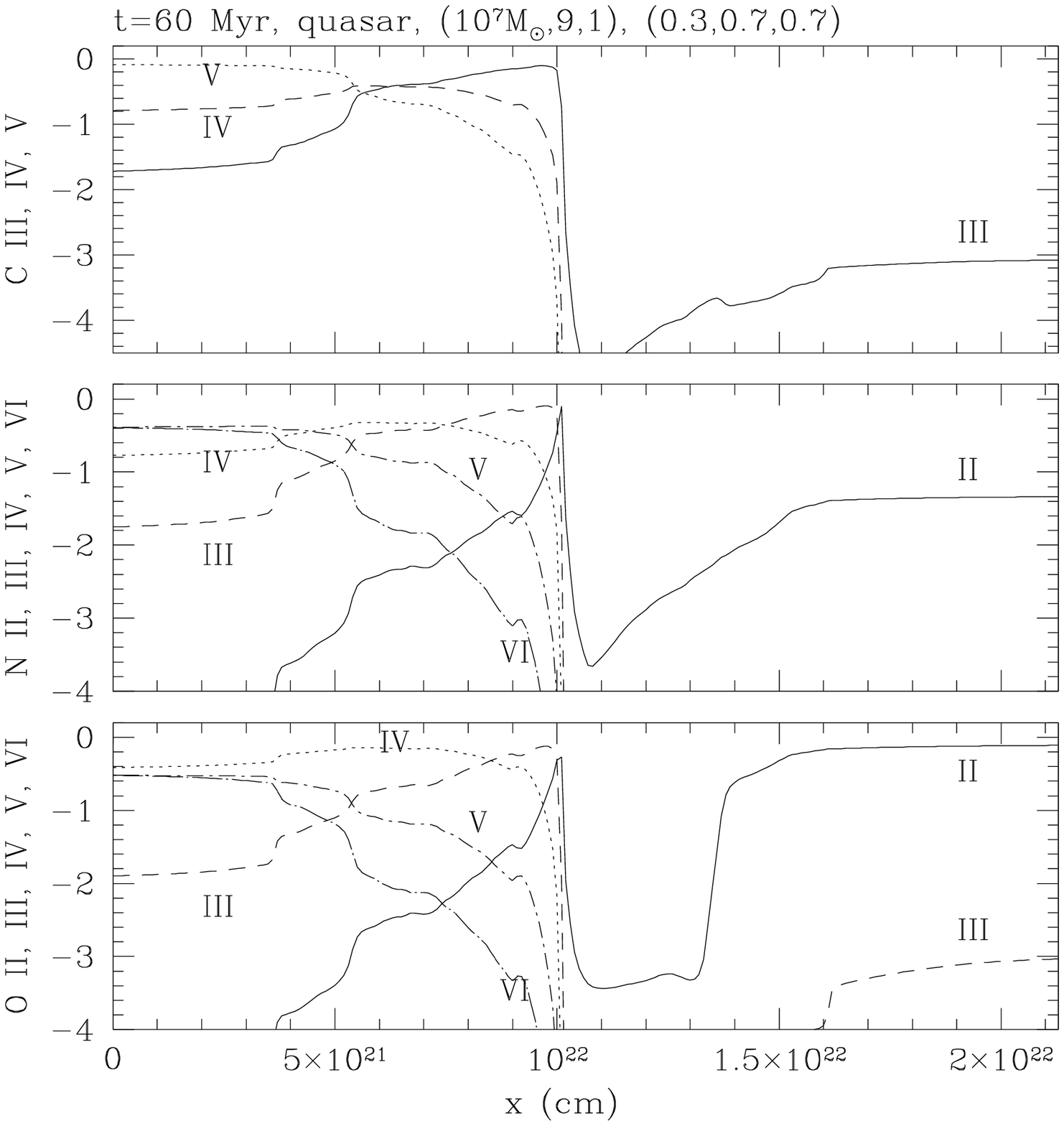,width=3.2in}}
\vspace{0pt}
\caption{{\bf Observational diagnostics I: ionization structure of metals.}
C, N, and O ionic fractions along symmetry
axis at $t=60\rm\,Myr$, for photoevaporating minihalo of
Figs.~5, 6. (a) (left)  STELLAR CASE; (b) (right) QUASAR CASE.}
\label{fig8}
\end{figure}

The results of our simulations on an $(r,x)$-grid with 
$256\times512$ cells are illustrated by Figures~5--9. 
The minihalo shields itself against ionizing photons,
traps the R-type I-front which enters the halo, causing it to decelerate
inside the halo to close to the sound speed of the ionized gas and
transform itself into a D-type
front, preceded by a shock. 
The side facing the source
expels a supersonic wind backwards towards the source, which shocks
the IGM outside the minihalo. The wind grows more isotropic with time as
the remaining neutral halo material
is photoevaporated. Since this gas was initially bound to a dark halo with 
$\sigma_V<\rm 10\,km\,s^{-1}$, photoevaporation proceeds unimpeded by
gravity.
Figures~5 and 6
show the structure of the photoevaporative flow 60~Myrs after the
global I-front first overtakes the minihalo, with key features of
the flow indicated by the labels on the temperature plot in Figure~6.
Figure~7(a) shows how the neutral mass of the gas initially
within the original hydrostatic sphere gradually declines as
the minihalo photoevaporates, within $t_{\rm ev}\approx250$ (100)~Myrs for the
stellar (quasar) cases, respectively.
The gradual decay of the opaque cross section of the
minihalo as seen by the source is illustrated by Figure~7(b).

\begin{figure}[b!] 
\vspace{-30pt}
\centerline{\epsfig{file=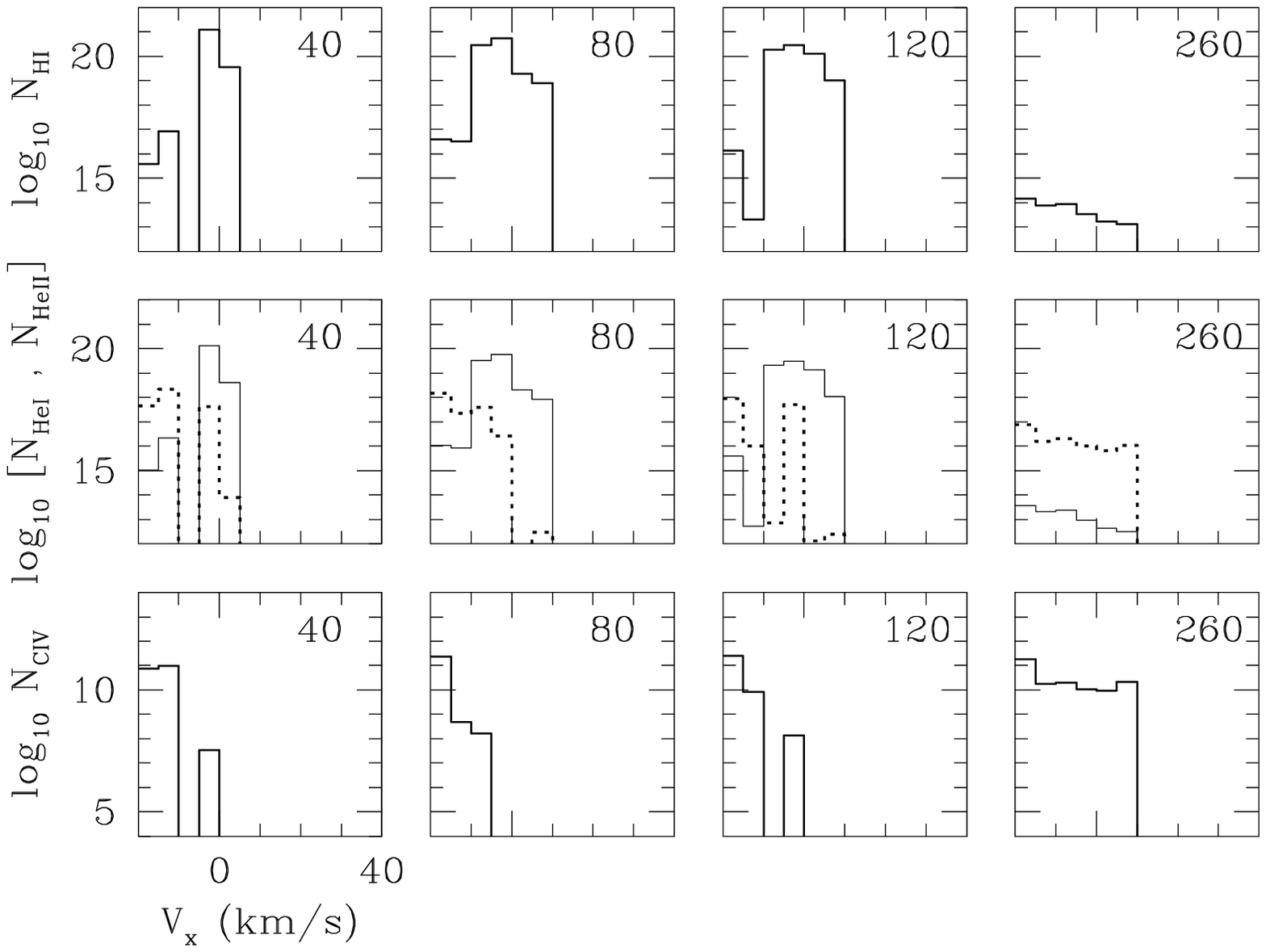,width=3.6in}\hskip-0.7in
            \epsfig{file=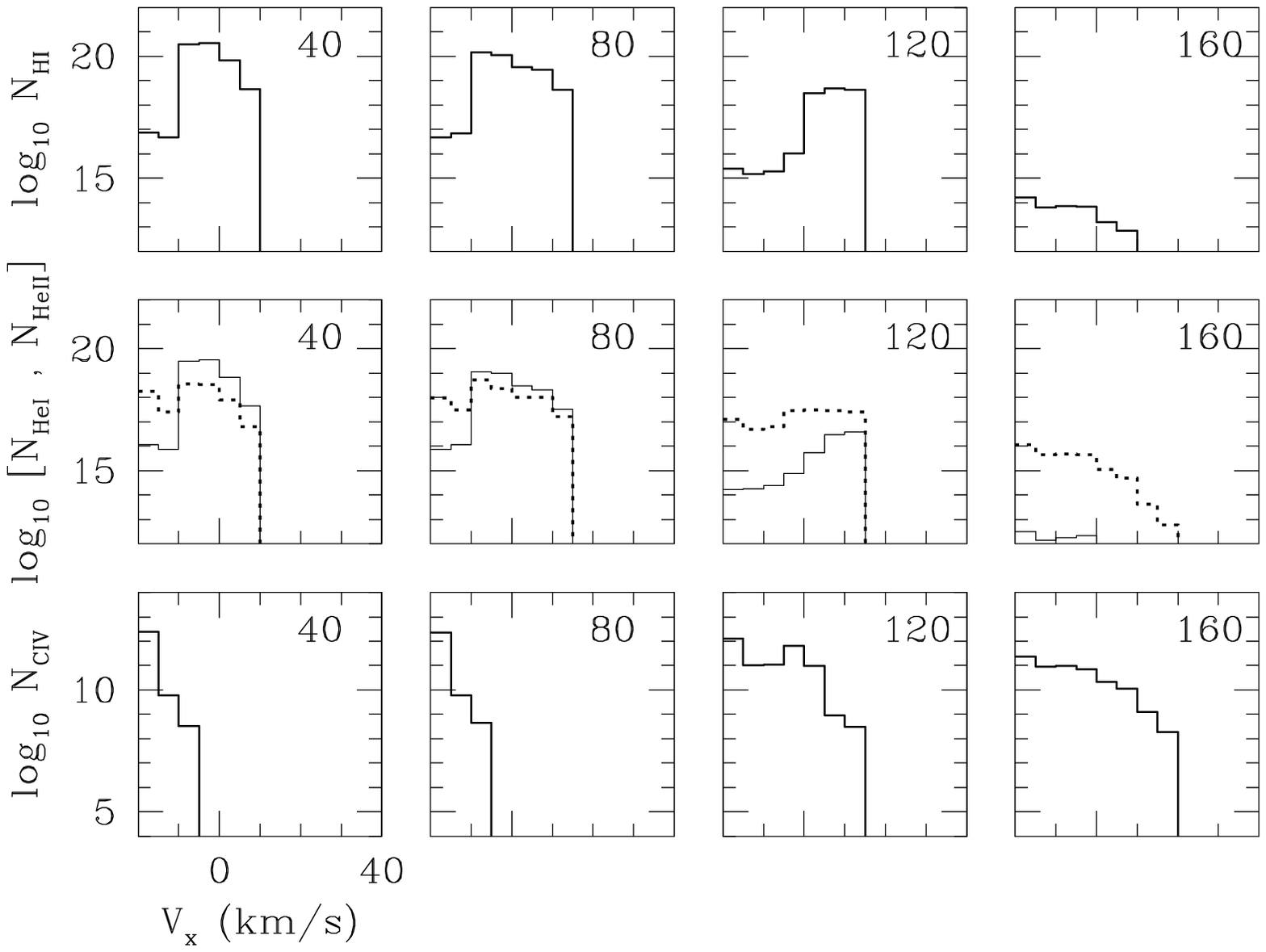,width=3.6in}}
\vspace{-60pt}
\caption{{\bf Observational diagnostics II. Absorption lines.}
Minihalo column densities ($\rm cm^{-2}$) along symmetry axis for gas at
different velocities, for photoevaporating minihalo of Figs.~5, 6. 
(a) (left)  STELLAR CASE; (b) (right) QUASAR CASE.
(Top) H~I; (Middle) He~I (solid) and
He~II (dotted); (Bottom) C~IV (i.e. if $\rm[C]/[C]_\odot=\xi\times10^{-3}$,
then plotted values are $N_{\rm CIV}/\xi$).
Each box labelled with time (in Myrs) since  arrival of intergalactic
I-front. }
\label{fig9}
\end{figure}

Some observational signatures of this process are shown in Figures~8 and~9.
Figure~8 shows the spatial variation of the relative abundances of
C, N, and O ions along the symmetry axis after 60~Myrs.
While the quasar case shows the presence at 60~Myrs
of low as well as high ionization stages for the metals, the softer spectrum
of the stellar case yields less highly ionized gas on the ionized side of 
the I-front (e.g. mostly C~III, N~III, O~III) and the neutral side
as well (e.g. C~II, N~I, O~I and II).
The column densities of H~I, He~I and II, and C~IV for minihalo gas of
different velocities as seen along the symmetry axis at different times
are shown in Figure~9. 
At early times, the minihalo gas resembles a weak
Damped Ly$\alpha$ (``DLA'') absorber with small velocity width 
($\gtrsim10\rm\,km\,s^{-1}$) and $N_{\rm H\,I}\gtrsim10^{20}\rm cm^{-2}$,
with a Ly$\alpha$-Forest(``LF'')-like red wing 
($\hbox{velocity width}\,\gtrsim10\,\rm km\,s^{-1}$)
with $N_{\rm H\,I}\gtrsim10^{16}\rm cm^{-2}$ on the side moving toward
the source, with a He~I profile which mimics that of H~I but with 
$N_{\rm He\,I}/N_{\rm H\, I}\sim {\rm [He]/[H]}$, and with a weak C~IV 
feature with $N_{\rm C\,IV}\sim10^{11}\,(10^{12})\xi\,\rm cm^{-2}$ for the
stellar (quasar) cases, respectively, displaced in
this same asymmetric way from the velocity of peak H~I column
density, where $\xi\equiv[{\rm C}]/[{\rm C}]_\odot\times10^3$.
For He~II at early times, the stellar case has 
$N_{\rm He\,II}\approx10^{18}\rm cm^{-2}$ shifted by 10's of 
$\rm km/sec$ to the red of the H~I peak, while for the quasar case,
He~II simply follows the H~I profile, except that 
$N_{\rm He\,II}/N_{\rm H\,I}\approx10$ in the red wing but 
$N_{\rm He\,II}/N_{\rm H\,I}\approx10^{-2}$ in the central H~I 
feature. After 260 (160)~Myr,
however, only a LF-like H~I feature with column density
$N_{\rm H\,I}\sim10^{14}\,\rm cm^{-2}$ remains, with 
$N_{\rm He\,I}/N_{\rm H\, I}\sim 1/4\, (\lesssim1/30)$, 
$N_{\rm He\,II}/N_{\rm H\,I}\sim10^3\, (10^2)$, and 
$N_{\rm C\,IV}/N_{\rm H\,I}\sim3(1)\times\rm[C]/[C]_\odot$ 
for the stellar (quasar) cases, respectively.

As described above, intervening minihalos like these are expected to be
ubiquitous along the line of sight to high redshift sources.
With photoevaporation times $t_{\rm ev}\gtrsim100\,\rm Myr$, this
process can continue down to redshifts significantly below $z=10$.
For stellar sources in the $\Lambda$CDM model, these simulations
show that photoevaporation of a $10^7M_\odot$ minihalo
which begins at $z_{\rm initial}=9$ can take 250 Myr to finish,
at $z_{\rm final}=6.8$, during which time such minihalos can survive without
merging into larger halos.
Observations of the absorption spectra of high redshift sources like those 
which reionized the universe should reveal the presence of 
these photoevaporative 
flows and provide a useful diagnostic of the reionization process.

{\bf Dwarf Galaxy Suppression and Reionization.}
As emphasized by \cite{sgb94}, the reheating of the IGM which
accompanied its reionization must have filtered the
linear growth of baryonic fluctuations in the IGM, 
thereby reducing the baryon
collapsed fraction and preventing baryons from condensing out
further into the minihalos (see also \cite{co93,co00,gnedin00b,gh98,vs99}).
A related effect, the suppression of baryon accretion onto dark matter halos,
has also been studied, by 1D \cite{htl96,ki00,ktusi00,tw96} 
and 3D simulations (e.g. \cite{ns97,qke96,whk97}).
The current conclusion is expressed in terms of the threshold 
circular velocity, $v_c$, below which baryonic infall
and star formation were suppressed by photoionization:
$v_c\sim30\,\rm km\,s^{-1}$ for complete suppression, 
partial suppression of infall found 
to extend even to $v_c\sim75\,\rm km\,s^{-1}$.
This suppression of gas accretion onto low-mass halos by the
feedback effect of reionization may naturally explain why
there are so many fewer dwarf galaxies observed in the Local Group
than are predicted by N-body simulations of the CDM model \cite{bkw00}. 

{\bf Acknowledgments:}
I thank A. Raga, I. Iliev, and H. Martel for
their collaboration on the work presented here,
supported by grants NASA ATP NAG5-7363 and 
NAG5-7821, NSF ASC-9504046, and Texas 
Advanced Research Program 3658-0624-1999.

\end{document}